\documentclass[twocolumn,prb]{revtex4}
\usepackage{bm}
\usepackage{epsfig}
\newcommand{\nix}[1]{}

\begin{document}
\title{Interface-Induced Electron Spin Splitting  in  SiGe Heterostructures}

\author{L.E.~Golub}\email{golub@coherent.ioffe.ru} 
\author{E.L.~Ivchenko} 

\affiliation{A.F.~Ioffe Physico-Technical Institute, Russian Academy of Sciences, 194021 St.~Petersburg, Russia}

\begin{abstract}
Spin splitting of conduction electron states has been analyzed for
all possible point symmetries of SiGe quantum well structures. A
particular attention is paid to removal of spin degeneracy caused
by the rotoinversion asymmetry of a (001) heterointerface between
two diamond-lattice materials. Consequences of the spin splitting
on the electron spin relaxation time is discussed.
\end{abstract}

\maketitle

\section{Introduction}
Spin properties attract the great attention in recent years due to
attempts to realize an electronic device based on the spin of
carriers. Conduction electrons are obvious candidates for such
devices, particularly in nanostructures where electron energy
spectrum and shape of the envelope functions can be effectively
engineered by the growth design, application of electric or
magnetic fields as well as by illumination with light.

Various semiconductor materials are being involved in the
spintronics activities. SiGe quantum well structures are among
them.~\cite{ganichev,jantsch} Although bulk Si and Ge have a
center of symmetry, quantum-well structures grown from these
materials can lack such a center and allow the spin splitting of
the electronic subbands.

The quantum engineering of spintronics devices is usually focused
on the Rashba spin-dependent contribution to the electron
effective Hamiltonian in heterostructures. This contribution
appears due asymmetry of the heterostructure (the so-called
structure inversion asymmetry, or SIA) and has no relation to the
properties of a bulk semiconductor. It is commonly believed that
the Dresselhaus contribution caused by bulk inversion asymmetry
(BIA) is absent in structures grown from centrosymmetric
materials. In the present work we show for the first time that the
Dresselhaus-like spin splitting  is possible in heterostructures
made of Si and Ge, as an alternative to the Rashba effect, and it
occurs due to the anisotropy of chemical bonds at interfaces.

\section{Symmetry of S\lowercase{i}/G\lowercase{e} quantum wells
and linear-in-${\bm k}$ spin splitting} \label{point_symmetry}

We consider quantum well (QW) structures grown along the axis $z
\parallel [001]$, and introduce two orthogonal directions in the
interface plane, $x \parallel [100]$ and $y \parallel [010]$.
Depending on the properties of an interface between
Si$_{1-x}$Ge$_x$ and Si, its symmetry on average can be C$_{\rm
2v}$ or $C_{\rm 4v}$.~\cite{xiao} The former point group describes
the symmetry of an ideal heterointerface with the interfacial
chemical bonds lying in the same plane. A nonideal interface
containing monoatomic fluctuations has two kinds of flat areas
with interfacial planes shifted with respect to each other by a
quarter of the lattice constant. The local symmetry of each
area is C$_{\rm 2v}$ as well. However if the both kinds are
equally distributed, the interface overall symmetry increases up
to $C_{\rm 4v}$. It follows then that the symmetry of a
Si$_{1-x}$Ge$_x$/Si quantum well structure containing two
interfaces is described by one of five point groups: $D_{\rm 2d}$
or $D_{\rm 2h}$ in case of two ideal interfaces with odd or even
number of monolayers between them; $C_{\rm 2v}$ for a pair of
ideal and rough interfaces; $C_{\rm 4v}$ or $D_{\rm 4h}$ for two
non-ideal interfaces of the overall symmetry $C_{\rm 4v}$ each |
see Fig.~1.

\begin{figure*}
\epsfxsize=\textwidth \epsfysize=0.3\textwidth
\centering{\epsfbox{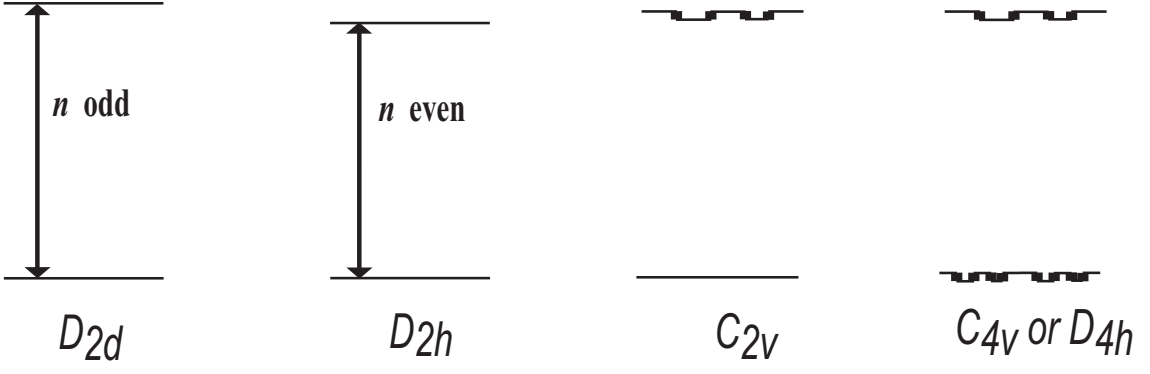}} \caption{Different interface
profiles and QW point symmetries. $n$ is a number of monoatomic layers.}
\end{figure*}

The lowest conduction band in (001)-SiGe/Si QWs is located near
the $X$ point of the Brillouin zone. At this point, the bulk
electron states form projective representations of the point group
$D_{\rm 4h}$. All five above-mentioned groups are subgroups of
$D_{\rm 4h}$. Two of them, $D_{\rm 4h}$ and $D_{\rm 2h}$, contains
the space inversion operation and forbid the spin splitting of
electronic states. Three remaining groups allow the spin-dependent
linear-in-${\bm k}$ term ${\cal H}_{{\bm k}} = \gamma_{\alpha \beta}
\sigma_{\alpha} k_{\beta}$. Here $\sigma_{\alpha}$ are the Pauli
matrices and ${\bm k} = (k_x, k_y)$ is the in-plane electron wave
vector. Note that in the following we consider the electronic
states attached to the $X_z$ valley which lies lower than $X_x$
and $X_y$ valleys.~\cite{aleshkin}

In the group $D_{\rm 2d}$, the Dresselhaus-like term $\sigma_x k_x -\sigma_y k_y$ is an only invariant which can be constructed from
the products $\sigma_{\alpha} k_{\beta}$. In particular, the
Rashba term $\sigma_x k_y - \sigma_y k_x$ is forbidden. On the
contrary, in the group $C_{\rm 4v}$, the only invariant
combination is $\sigma_x k_y - \sigma_y k_x$. The analysis shows
that both combinations $\sigma_x k_x - \sigma_y k_y$ and $\sigma_x
k_y - \sigma_y k_x$ are invariants of the group $C_{\rm 2v}$.
Hence the latter allows both Dresselhaus- and Rashba-like
spin-dependent terms.

\section{Spin splitting: microscopic model}\label{microscopic_X}
Microscopically, a linear-in-$\bm k$ correction to the
conduction-band effective Hamiltonian is given by the second order
of the perturbation theory
\begin{eqnarray}\label{second_order}
&& \Delta {\cal H}_{i,i'} =\\ 
&& \sum_{n} { \langle c i | H_{\rm SO} |  n \rangle \langle n | H_{\bm {k p}} | c i' \rangle + \langle
c i | H_{\bm {k p}} | n \rangle \langle n | H_{\rm SO} | c i'
\rangle \over E_c - E_v}\:.
\nonumber
\end{eqnarray}
Here $H_{\bm{k p}}$ and $H_{\rm SO}$ are the $\bm{k \cdot
\hat{p}}$ and spin-orbit interaction Hamiltonians, respectively.
We take into consideration only the coupling of the conduction
states $(c, X_1, i)$ with the valence states $n = (v, X_4, j)$,
where the indices $i,j$ enumerate the degenerate states. The
symmetry properties of the Bloch functions at the $X$ point of a
diamond-lattice semiconductor crystal coincide with those of the
following functions~\cite{yu/cardona}
\begin{equation}
\label{X1} X_1\mbox{-states}: \left\{
\begin{array}{cc}
S=\cos{\left( {2 \pi z / a}\right)} \\
Z=\sin{\left( {2 \pi z / a}\right)}
\end{array}
\right.,
\end{equation}
\begin{equation}
\label{X4} X_4\mbox{-states}: \left\{
\begin{array}{cc}
X = \sin{\left( {2 \pi x / a}\right)} \cos{\left( {2 \pi y /
a}\right)} \\
Y = \cos{\left( {2 \pi x / a}\right)} \sin{\left( {2 \pi y /
a}\right)}
\end{array}
\right. ,
\end{equation}
where $a$ is the lattice constant. For the bulk states $X_1, X_4$ in the bases~(\ref{X1}),~(\ref{X4}), the interband matrix elements can be presented as
\begin{equation} \label{kpso}
H_{\bm{k p}} =  {\cal P} \left( \begin{array}{cc}
k_x & k_y\\
-k_y & -k_x \end{array} \right), \:\:\:\:
H_{\rm SO} = {\cal U} \left( \begin{array}{cc} \sigma_x & -\sigma_y\\
-\sigma_y & \sigma_x
\end{array}
\right)\:.
\end{equation}
Here ${\cal P} = (\hbar/m_0) \left< S | \hat{p}_x | X \right>$,
${\cal U} = \left< S | U_x | X \right>$, $\hat{\bm p}$ is the
momentum operator, and $U_x$ is the $x$-component of the
pseudovector $\vec{U}$ which enters into the spin-orbit
Hamiltonian $H_{\rm SO} = \vec{\sigma} \cdot \vec{U}$.

The substitution of (\ref{kpso}) into (\ref{second_order}) gives a
spin-dependent matrix $\Delta {\cal H}$ proportional to
\[
(\sigma_x k_x - \sigma_y k_y) \left( \begin{array}{cc} 1 & 0\\
 0 & -1 \end{array} \right)
\]
which however {\em does not} lead to removal of the degeneracy of
the $X_1$-states in the bulk centrosymmetric material as expected.
The splitting can be achieved if one takes into account the
anisotropy of chemical bonds at the interfaces which results in
$\delta$-functional contributions to $H_{\bm{k p}}$ and $H_{\rm
SO}$ of the form
\begin{equation} \label{deltakpso}
\Delta H_{\bm{k p}} = \hat{V}_{\bm{k p}}\: \delta(z - z_{if})\:,\:
\Delta H_{\rm SO} = \hat{V}_{\rm SO}\: \delta(z - z_{if})\:,
\end{equation}
where $z_{if}$ is the interface coordinate, and the matrices
$\hat{V}_{\bm{k p}}, \hat{V}_{\rm SO}$ have few linearly
independent components. In the case of the lowest symmetry under
study, $C_{\rm 2v}$, each matrix is determined by four
linearly-independent parameters
\begin{eqnarray} \label{kp_C2v}
\hat{V}_{\bm{k p}} &=& \left(
\begin{array}{cc}
P_1 k_x + P_2 k_y & P_1 k_y + P_2 k_x\\ P_3 k_y + P_4 k_x & P_3
k_x + P_4 k_y\end{array} \right)\:, \\
 \label{so_C2v}
\hat{V}_{\rm SO} &=& \left(
\begin{array}{cc}
U_1 \sigma_x - U_2 \sigma_y & -U_1 \sigma_y + U_2 \sigma_x\\ -U_3
\sigma_y + U_4 \sigma_x & U_3 \sigma_x - U_4 \sigma_y\end{array}
\right)\:.
\end{eqnarray}
In the bases~(\ref{X1}),~(\ref{X4}),  the parameters $P_l$ and $U_l$ $(l=1
\div  4)$ are purely imaginary.

According to Eq.~(\ref{second_order}), the linear in $k_{x,y}$
correction to the conduction-band Hamiltonian for a single
interface of the $C_{\rm 2v}$-symmetry, is given by
\begin{equation} \label{c2vif}
\Delta {\cal H}_{C_{\rm 2v}} = \left( \begin{array}{cc} 1 & 0\\
 0 & 1 \end{array} \right) H_{if} \delta(z - z_{if}) \:,
\end{equation}
where $H_{if}$ is a linear combination of the Pauli matrices.
Treating the perturbation (\ref{deltakpso}) in the first-order
approximation, one can obtain
\begin{eqnarray}
\label{H_C2v} 
&&H_{if} = \\
&-& {{\cal P} \over E_0} \biggl[(\sigma_x k_x - \sigma_y k_y) (U_1 - U_3)  + (\sigma_x k_y - \sigma_y k_x) (U_2 - U_4) \biggr] 
\nonumber\\
&-& {{\cal U} \over E_0} \biggl[ (\sigma_x k_x - \sigma_y k_y) (P_1 + P_3) + (\sigma_x k_y - \sigma_y k_x) (P_2 + P_4) \biggr],
\nonumber
\end{eqnarray}
where $E_0$ is the energy separation between $X_1$ and $X_4$
states.

The electron effective Hamiltonian in an ideal QW contains a sum
of two contributions (\ref{c2vif}) related to the left- and
right-hand-side interfaces. The corresponding parameters $U_l^L,
U_l^R$ or $P_l^L, P^R_l$ are interconnected by
\begin{equation}\label{U_D2h}
[U_1, U_2, U_3, U_4]_R = [U_3, U_4, U_1, U_2]_L,
\end{equation}
\[
[P_1, P_2, P_3, P_4]_R = [- P_3, - P_4, - P_1, -P_2]_L,
\]
if the QW contains an even number of monoatomic layers ($D_{\rm
2h}$ symmetry), and
\begin{equation} \label{U_D2d}
[U_1, U_2, U_3, U_4]_R = [U_1, - U_2, U_3, - U_4]_L,
\end{equation}
\[
[P_1, P_2, P_3, P_4]_R = [P_1, -P_2, P_3, -P_4]_L,
\]
if the QW contains an odd number of monoatomic layers ($D_{\rm
2d}$ symmetry). It readily follows from here that, in the former
case, the electronic states in Si/Ge QWs are spin-degenerate
($H_{if}=0$) and, in the latter, the spin degeneracy is removed by
a Dresselhaus-like linear-in-${\bm k}$ terms. The similar analysis
can be carried out for non-ideal Si/Ge QW structures of the
$C_{\rm 2v}$ and $C_{\rm 4v}$ symmetry.

\section{Conclusion}
In this work, we have analyzed spin splitting of electron states
in Si/Ge heterostructures of all possible symmetries. This opens a
possibility to discuss the recent spin relaxation times
measurements in these heterostructures.~\cite{jantsch} The work on
estimation of the interface parameters $U_l$ and $P_l$ in the
microscopic tight-binding model is in progress.


\section*{Acknowledgements} This work is financially supported by the
RFBR, DFG, INTAS, by the Programmes of Russian Ministry of Science
and Presidium of RAS, and by the Dynasty Foundation and ICFPM.

\end{document}